\documentclass[a4paper,twocolumn]{article}
\usepackage{graphicx}
\usepackage{amsmath}

\begin{document}

\title{Perturbational approach to the possible quantum capacity of additive Gaussian
quantum channel}
\author{Xiao-yu Chen \\
{\small {College of Information and Electronic Engineering, Zhejiang
Gongshang University, Hangzhou, 310018, China}}}
\date{}
\maketitle

\begin{abstract}
For a quantum channel with additive Gaussian quantum noise, at the large
input energy side, we prove that the one shot capacity is achieved by the
thermal noise state for all Gaussian state inputs. For a general case of $n$
copies input, we show that up to first order perturbation, any non-Gaussian
perturbation to the product of identical thermal states input has a less
quantum information transmission rate when the input energy tends to
infinitive.

PACS number(s): 03.67.-a, 42.50.Dv, 89.70.+c
\end{abstract}
\\

Quantum capacity is one of the main issues in quantum information theory. It
is concerned with the transmission ability of unknown quantum state on a
given quantum channel. The critical quantity involved in the quantum
capacity is the coherent information (CI) $I_c(\sigma ,\mathcal{E})=S(%
\mathcal{E}(\sigma ))-S(\sigma ^{QR^{\prime }})$ \cite{Schumacher} \cite
{Lloyd}. Here $S(\varrho )=-$Tr$\varrho \log _2\varrho $ is the von Neumann
entropy, $\sigma $ is the input state, the application of the channel $%
\mathcal{E}$ results the output state $\mathcal{E}(\sigma )$; $\sigma
^{QR^{\prime }}=$ $(\mathcal{E}\otimes \mathbf{I})(\left| \psi \right\rangle
\left\langle \psi \right| )$, with $R$ referred to the 'reference' system%
\cite{Schumacher} (the system under process is $Q$ system with annihilation
and creation operators $a$ and $a^{\dagger }$, we denote $\sigma ^Q$ as $%
\sigma $ for simplicity), $\left| \psi \right\rangle $ is the purification
of $\sigma $. The quantum channel capacity is\cite{Devetak}\cite{Barnum}\cite
{Horodecki}
\begin{equation}
Q=\lim_{n\rightarrow \infty }\sup_{\sigma _n}\frac 1nI_c(\sigma _n,\mathcal{E%
}^{\otimes n}).  \label{wave0}
\end{equation}
Quantum capacity exhibits a kind of nonadditivity \cite{DiVincenzo} that
makes it extremely hard to deal with. The first example with calculable
quantum capacity is quantum erasure channel\cite{Bennett}. Other examples
are dephasing qubit channel\cite{Devetak0}, amplitude damping qubit channel%
\cite{Giovannetti}, and continuous variable lossy channel\cite{Wolf}, where
the channels are either degradable or anti-degradable\cite{Caruso}. Gaussian
quantum channel \cite{Harrington} (additive classical Gaussian channel
followed Holevo \cite{Holevo} ) is quite essential in quantum information
theory. Unfortunately, this channel is neither degradable nor anti-degradable%
\cite{Holevo0} \cite{Caruso0} makes the technics developed for calculating
the quantum capacity unapplicable.

The quantum capacity of the Gaussian quantum channel has been conjectured as%
\cite{Holevo}
\begin{equation}
Q=\max \{0,-\log _2(eN_n)\},  \label{wave1}
\end{equation}
where $N_n$ specifies the Gaussian quantum channel. It can be achieved by
quantum error-correction codes\cite{Harrington}. For additive Gaussian
quantum channel, we have \cite{Harrington} \cite{Chen}
\begin{equation}
\mathcal{E}(\sigma )=\frac 1{N_n}\int \frac{d^2\alpha }\pi \exp (-\left|
\alpha \right| ^2/N_n)\mathcal{D}\left( \alpha \right) \sigma \mathcal{D}%
^{\dagger }(\alpha ),  \label{wave2}
\end{equation}
where $\mathcal{D}\left( \alpha \right) =$ $\exp (\alpha a^{\dagger }-\alpha
^{*}a)$ is the displacement operator. Any quantum state $\sigma $ can be
equivalently specified by its characteristic function $\chi _\sigma (\mu
)=Tr[\sigma \mathcal{D}(\mu )]$, and inversely $\mathcal{\sigma }=\int
[\prod_i\frac{d^2\mu _i}\pi ]\chi _{\mathcal{\sigma }}(\mu )\mathcal{D}(-\mu
)$. The characteristic function of the noisy state $\sigma ^{\prime }=%
\mathcal{E}(\sigma )$ is $\chi _\sigma ^{\prime }(\mu )=\chi _\sigma (\mu
)e^{-N_n\left| \mu \right| ^2}$.

A single mode thermal state $\rho $ has a characteristic function of the
form $\chi (\mu )=\exp [-(N+\frac 12)\left| \mu \right| ^2]$, and we have $%
\rho =\int \frac{d^2\mu }\pi \chi (\mu )\mathcal{D}(-\mu )=$ $%
(1-v)v^{a^{\dagger }a}$ with $v=N/(N+1),$[conventionally in the following, $%
v_x=N_x/(N_x+1)$], where $N$ is the average photon number. The noisy state
is $\rho ^{\prime }=\mathcal{E}(\rho )=(1-v^{\prime })v^{\prime a^{\dagger
}a},$ with average photon number $N^{\prime }=N+N_n,$ this accounts for the
'additive' of the channel.

We now consider a Gaussian state $\rho _G$ (which comprises thermal noise
state as its special case) input to the channel. A single mode Gaussian
state is described by its real correlation matrix $\alpha $ (we drop the
first moments of the state for they can be removed by local operations). We
have $\alpha =\left[
\begin{array}{ll}
\alpha _{qq} & \alpha _{qp} \\
\alpha _{qp} & \alpha _{pp}
\end{array}
\right] $ . The energy of the Gaussian state is $E=Tr[(a^{\dagger }a+\frac
12)\rho _G]=\frac 12(\alpha _{qq}+\alpha _{pp}).$ For a Gaussian state input
$\rho _G$, the output $\rho _G^{\prime }$ and the joint output state $\rho
_G^{QR^{\prime }}$ are still Gaussian. The symplectic eigenvalues \cite
{Holevo} of these states can be obtained. The coherent information is
\begin{equation}
I_c(\rho _G,\mathcal{E})=g(d_0-\frac 12)-g(d_1-\frac 12)-g(d_2-\frac 12),
\end{equation}
with
\begin{eqnarray}
d_0 &=&\sqrt{N_n^2+2N_nE+E^2x} \\
d_{1,2} &=&\sqrt{\frac 12[N_n^2+2N_nE+\frac 12\pm N_nD_G]},
\end{eqnarray}
where $D_G=\sqrt{(N_n+2E)^2+1-(2E)^2x},$ with $x=\det (\alpha )/E^2$. Here $%
g(s)=(s+1)\log (s+1)-s\log s$ is the bosonic entropy function. When $E$ is
fixed, the maximum of $x$ can be obtained with the derivatives on $x-\lambda
(\alpha _{qq}+\alpha _{pp}-2E),$ where $\lambda $ is the Lagrange
multiplier. It follows that the maximum value $x=1$ is achieved when $\alpha
_{qq}=\alpha _{pp}=\sqrt{E},$ $\alpha _{qp}=0$. At sufficiently large input
energy $E$, calculating $\frac{dI_c(\rho _G,\mathcal{E})}{dx}$ and expanding
the expression with $E^{-1},$ we then obtain
\begin{equation}
\frac{dI_c(\rho _G,\mathcal{E})}{dx}=\frac 1{2Ex^2}(\frac
1{3N_n}-2N_n)+o(\frac 1{E^2}),
\end{equation}
Since $N_n<1/e$ (Otherwise $I_c=0$, this is shown by (\ref{wave1}), also by
numeric results of Gaussian input), thus $\frac{dI_c(\rho _G,\mathcal{E})}{dx%
}$ is positive for sufficiently large input energy. While $x$ has its global
maximum value $x=1,$ so the coherent information achieves its maximum at $%
x=1 $ which corresponds to thermal noise state input. Hence we can conclude
that for sufficient large but definite input energy, the one-shot quantum
information capacity of Gaussian quantum channel is achieved by thermal
noise state input of all Gaussian state inputs.

For thermal state input $\rho $, denote the annihilation and creation
operators of the 'reference' $R$ system as $b$ and $b^{\dagger }$, we have%
\cite{Chen}
\begin{eqnarray}
\rho ^{QR^{\prime }} &=&(1-v)(1-v_n)\exp [\sqrt{v}(1-v_n)a^{\dagger
}b^{\dagger }]  \nonumber \\
&&\times v_n^{a^{\dagger }a}(vv_n)^{b^{\dagger }b}\exp [\sqrt{v}(1-v_n)ab],
\label{wave3}
\end{eqnarray}
The state $\rho ^{QR^{\prime }}$ can be written as $S_2(r)(\rho _A\otimes
\rho _B)S_2^{\dagger }(r),$ $S_2(r)=\exp [r(a^{\dagger }b^{\dagger }-ab)]$
is the two-mode squeezing operator, the squeezing parameter $r$ is
determined by $\tanh 2r=2\sqrt{N(N+1)}/(2N+N_n+1).$ $\rho _A$ and $\rho _B$
are two thermal states with average photon numbers $N_A$ and $N_B,$
respectively, where $N_{A,B}=\frac 12(D\pm N_n-1)$ with $D=\sqrt{%
N_n^2+2(2N+1)N_n+1}$. The coherent information will be \cite{Holevo}\cite
{Chen}
\begin{equation}
I_c(\rho ,\mathcal{E})=g(N+N_n)-g(N_A)-g(N_B).  \label{wave4}
\end{equation}
One of the useful formula is
\begin{equation}
N=N_B\cosh ^2r+(N_A+1)\sinh ^2r,  \label{wave5}
\end{equation}

To treat with the non-Gaussian perturbation, we need the following lemmas:

\textit{Lemma 1: }$(\mathcal{E}\otimes \mathbf{I)}(a^{\dagger k}\rho
^{QR}a^m)\mathbf{=}v^{-(k+m)/2}b^k\rho ^{QR^{\prime }}b^{\dagger m}$.

\textit{proof: } With the characteristic function $\chi ^{QR}$of $\rho
^{QR}, $ the lhs can be written as $\frac 1{N_n}\int \frac{d^2\alpha }\pi
\frac{d^4\mu }{\pi ^2}$ $\exp [-\frac{\left| \alpha \right| ^2}{N_n}+\mu
_1\alpha ^{*}-\alpha \mu _1^{*}]$ $(a^{\dagger }-\alpha ^{*})^k\chi
^{QR}(\mu )$ $\mathcal{D}(-\mu )(a-\alpha )^m$. After the integral on $\mu
=(\mu _1,\mu _2) $, and a displacement on $\alpha $ (in the ordered operator
product)$:$ $\alpha \rightarrow \alpha +a;\alpha ^{*}\rightarrow \alpha
^{*}+a^{\dagger }, $ the lhs can be further written as
\begin{eqnarray}
\frac 1{N_n}\int \frac{d^2\alpha }\pi :\alpha ^{*}{}^k\exp [-\frac{\left|
\alpha \right| ^2}{v_n}+\alpha (\frac{a^{\dagger }}{N_n}+\sqrt{v}b)]
\nonumber \\
\cdot \exp [\alpha ^{*}(\frac a{N_n}+\sqrt{v}b^{\dagger })-\frac{a^{\dagger
}a}{N_n}-b^{\dagger }b]\alpha ^m :,
\end{eqnarray}
where the notation $:H:$ refers to that $H$ is in its ordered operator
product, that is, all creation operators are at the left of the annihilation
operators. The integral can be worked out with the formula $I=\frac 1K\int
\frac{d^2\alpha }\pi \exp [-\frac{\left| \alpha \right| ^2}K+\alpha \sigma
+\alpha ^{*}\tau ]=\exp (K\sigma \tau ),$ and its derivatives
\begin{equation}
\frac{\partial ^{k+m}I}{\partial \sigma ^m\partial \tau ^k}=\sum_{l=0}^{\min
\{k,m\}}\binom kl\binom mll!K^{k+m-l}\sigma ^{k-l}\tau ^{m-l}\exp [K\sigma
\tau ].  \label{wave6}
\end{equation}
We have
\begin{eqnarray}
(\mathcal{E}\otimes \mathbf{I)}(a^{\dagger k}\rho ^{QR}a^m)
=:\sum_{l=0}^{\min \{k,m\}}\binom kl\binom mll!v_n^{k+m-l}  \nonumber \\
(\frac{a^{\dagger }}{N_n}+\sqrt{v}b)^{k-l}(\frac a{N_n}+\sqrt{v}b^{\dagger
})^{m-l}\rho ^{QR^{\prime }}:.  \label{wave7}
\end{eqnarray}
Removing the ordered notation by moving all creation operators to the left
of $\rho ^{QR^{\prime }}$ and all annihilation operators to the right of $%
\rho ^{QR^{\prime }}$, and exchanging $b$ with $\rho ^{QR^{\prime }}$%
according to
\begin{equation}
\rho ^{QR^{\prime }}b=[b-\sqrt{v}(1-v_n)a^{\dagger }]/(vv_n)\rho
^{QR^{\prime }},  \label{wave8}
\end{equation}
exchanging $b^{\dagger }$ with $b$ and further with $\rho ^{QR^{\prime }}$,
where the formula $b^{\dagger m}b^n=\sum_{i=0}^{\min \{m,n\}}\binom mi\binom
nii!(-1)^ib^{n-i}b^{\dagger (m-i)}$ is used, after all the summation, the
lemma 1 is proved.

\textit{Lemma 2:} $(\mathcal{E}\otimes \mathbf{I)}(a^k\rho ^{QR}a^{\dagger
m})\mathbf{=}v^{(k+m)/2}b^{\dagger k}\rho ^{QR^{\prime }}b^m.$

\textit{proof: }The lhs is $\frac 1{N_n}\int \frac{d^2\alpha }\pi (a-\alpha
)^k$ $\{\int \frac{d^4\mu }{\pi ^2}\exp [-\frac{\left| \alpha \right| ^2}{N_n%
}+\mu _1\alpha ^{*}-\alpha \mu _1^{*}]$ $\chi ^{QR}(\mu )\mathcal{D}(-\mu
)\}(a^{\dagger }-\alpha ^{*})^m,$ after the integral on $\mu $, it is
\begin{eqnarray}
&&\frac 1{N_n}\int \frac{d^2\alpha }\pi \exp [-\frac{\left| \alpha \right| ^2%
}{N_n}](a-\alpha )^k  \nonumber \\
&&\exp [\alpha a^{\dagger }+\sqrt{v}(a^{\dagger }-\alpha ^{*})b^{\dagger
}]\left| 00\right\rangle \left\langle 00\right|  \nonumber \\
&&\exp [\alpha ^{*}a+\sqrt{v}(a-\alpha )b](a^{\dagger }-\alpha ^{*})^m,
\end{eqnarray}
where the formula $:e^{-a^{\dagger }a}:=\left| 0\right\rangle \left\langle
0\right| $ is used. Note that $\left( a-\alpha \right) e^{\alpha a^{\dagger
}}=e^{\alpha a^{\dagger }}a,$ thus $(a-\alpha )^k\exp [\alpha a^{\dagger }+%
\sqrt{v}(a^{\dagger }-\alpha ^{*})b^{\dagger }]\left| 00\right\rangle =\exp
[\alpha a^{\dagger }+\sqrt{v}(a^{\dagger }-\alpha ^{*})b^{\dagger }](\sqrt{v}%
b^{\dagger })^k\left| 00\right\rangle .$ After the integral on $\alpha ,$
the lemma 2 is proved.

In the single mode situation, we expand the input state $\rho _\varepsilon $
at the vicinity of $\rho $, the characteristic function of the input state
is $\chi _\varepsilon (\mu )=$Tr$(\rho _\varepsilon D(\mu ))=\chi (\mu
)(1+\varepsilon f(\mu ,\mu ^{*})).$ The perturbation item $f(\mu ,\mu ^{*})$
is a polynomial of $\mu $ and $\mu ^{*}$. Typically, this may contain (1) $%
\left| \mu \right| ^{2n}$ ($n>1,$ the $n=1$ is a Gaussian type perturbation,
here we discuss single mode situation, thus $n$ can not be confused with
that appeared in (\ref{wave0}) where it stands for the number of the modes )
and (2) $c\mu ^n(-\mu ^{*})^l+c^{*}\mu ^l(-\mu )^{*n}$ $(n\neq l)$. The
first type of perturbation will contribute to the first order perturbation
of the eigenvalues of $\rho $, while the second type of perturbation has not
a first order perturbation to the eigenvalues of $\rho $, it can only
contribute to the second order perturbation.

For the first type perturbation $\left| \mu \right| ^{2n},$ We have $\chi
_\varepsilon (\mu )=\chi (\mu )(1+\varepsilon \left| \mu \right| ^{2n}),$
thus $\rho _\varepsilon =\int \exp [-(N+\frac 12)\left| \mu \right|
^2](1+\varepsilon \left| \mu \right| ^{2n})\mathcal{D}(-\mu )\frac{d^2\mu }%
\pi =$ $(1+\varepsilon (-1)^n\frac{d^n}{dN^n})\rho =$ $\rho +\varepsilon
\phi .$ The strict eigenvalues of $\rho _\varepsilon $ are $\lambda
_k^\varepsilon =\lambda _k+\varepsilon \phi _k,$ with $\lambda _k=(1-v)v^k$
and $\phi _k=\lambda _k\xi _k,$where $\xi _k=(1-v)^n\sum_{j=0}^{\min
\{n,k\}}(-1)^jn!\binom nj\binom kjN^{-j}.$ $\xi _k$ is the eigenvalue of an
operator
\begin{equation}
\xi (a,a^{\dagger })=(1-v)^n\sum_{j=0}^n(-1)^j\frac{n!}{j!}\binom
njN^{-j}a^{\dagger j}a^j,  \label{wave10}
\end{equation}
with its eigenvector $\left| k\right\rangle .$ The moments of $\phi $ can be
calculated by $Tr(a^{\dagger l}a^l\phi )=(-1)^n\frac{d^n}{dN^n}Tr(a^{\dagger
l}a^l\rho )=(-1)^n$ $\frac{d^n}{dN^n}(l!N^l)$, thus for $l<n$, the moments
are nullified. The entropy of $\rho _\varepsilon $ can be expanded up to the
second order of $\varepsilon $ as $S(\rho _\varepsilon )=S(\rho )-\frac
12\varepsilon ^2\sum_k\phi _k^2/\lambda _k+o(\varepsilon ^3),$ where $%
Tr(\phi )=0$ and $Tr(a^{\dagger }a\phi )=0$ are used. We have $\sum_k\phi
_k^2/\lambda _k=Tr(\xi \phi )=$ $(1-v)^n\sum_{j=0}^n(-1)^j\frac{n!}{j!}%
\binom njN^{-j}Tr(a^{\dagger j}a^j\phi )=$ $(1-v)^nN^{-n}\left( n!\right)
^2. $ The calculation of the entropy of the noisy state $\rho _e^{\prime }$
is straightforward, it is
\begin{equation}
S(\rho _\varepsilon ^{\prime })=S(\rho ^{\prime })-\frac 12\varepsilon ^2%
\frac{\left( n!\right) ^2}{N^{\prime n}(N^{\prime }+1)^n}+o(\varepsilon ^3).
\end{equation}
The purification of $\rho $ is $\rho ^{QR}=\sum_{km}\sqrt{\lambda _k\lambda
_m}\left| kk\right\rangle \left\langle mm\right| .$ The state $\rho
_\varepsilon ^{QR}$ then is expanded in $\varepsilon $ to the linear item
(the $\varepsilon ^2$ term is less important in the large $N$ limit). $\rho
_\varepsilon ^{QR}=\rho ^{QR}+\varepsilon \Phi ,$ with $\Phi =\frac 12(\Phi
_0+\Phi _0^{\dagger }),$ $\Phi _0=(1-v)^n\sum_{j=0}^n(-1)^j\frac{n!}{j!}%
\binom njN^{-j}a^{\dagger j}b^{\dagger j}v^{j/2}\rho ^{QR}.$ Using lemma 1,
we arrive at
\begin{equation}
(\mathcal{E}\otimes \mathbf{I)}\rho _\varepsilon ^{QR}=\rho ^{QR^{\prime
}}+\varepsilon \Phi ^{\prime },
\end{equation}
with $\Phi ^{\prime }=\frac 12(\Phi _0^{\prime }+\Phi _0^{\prime \dagger }),$
$\Phi _0^{\prime }=\xi (b,b^{\dagger })\rho ^{QR^{\prime }}.$ The
eigenstates of $\rho ^{QR^{\prime }}$ are $\left| km\right\rangle ^{\prime
}= $ $S_2(r)\left| km\right\rangle $ with eigenvalues $\lambda
_{km}=(1-v_A)v_A^k$ $(1-v_B)v_B^m.$ The corresponding annihilation operators
are transformed to $a^{\prime }=S_2(r)aS_2^{\dagger }(r)=a\cosh r-b^{\dagger
}\sinh r,$ $b^{\prime }=S_2(r)bS_2^{\dagger }(r)=b\cosh r-a^{\dagger }\sinh
r.$ The first order perturbation to the eigenvalues will be $\Phi
_{km}^{\prime }=\left\langle km\right| ^{\prime }\Phi ^{\prime }\left|
km\right\rangle ^{\prime }=\lambda _{km}\left\langle km\right| ^{\prime }\xi
(b,b^{\dagger })\left| km\right\rangle ^{\prime }$. Note that
\begin{eqnarray}
\left\langle km\right| ^{\prime }b^{\dagger j}b^j\left| km\right\rangle
^{\prime } &=&\left\langle km\right| ^{\prime }\sum_{i=0}^j\binom
ji^2b^{\prime \dagger i}b^{\prime i}a^{\prime (j-i)}a^{\prime \dagger (j-i)}
\nonumber \\
&&\times \cosh ^{2i}r\sinh ^{2(j-i)}r\left| km\right\rangle ^{\prime }
\end{eqnarray}
We can construct an operator $\Omega $ which is the diagonal part of $\Phi
^{\prime }$ in the basis $\left| km\right\rangle ^{\prime }$, that is,
\begin{eqnarray}
\Omega &=&N^{-n}(N+1)^{-n}\sum_{j=0}^n\binom nj^2[N_B(N_B+1)\cosh ^2r]^j
\nonumber \\
&&[N_A(N_A+1)\sinh ^2r]^{n-j}\frac{\partial ^n}{\partial N_B^j\partial
N_A^{n-j}}\rho ^{QR^{\prime }}.  \label{wave11}
\end{eqnarray}
Where we have used the following lemma.

\textit{Lemma 3:} 1. $a^{\dagger j}a^j\rho =N^j\sum_{i=0}^j\frac{j!}{i!}%
\binom ji(N+1)^i\frac{d^i\rho }{dN^i};$ 2. $a^ja^{\dagger j}\rho
=(N+1)^j\sum_{i=0}^j\frac{j!}{i!}\binom jiN^i\frac{d^i\rho }{dN^{^i}}.$

\textit{Proof: }These two equalities can be proved with mathematical
induction.

For $\rho ^{QR^{\prime }}$ is a direct product state in the basis $\left|
km\right\rangle ^{\prime },$ we can treat the two new modes separately. Up
to $\varepsilon ^2$ item, the entropy will be $S(\rho _\varepsilon
^{QR^{\prime }})=S(\rho ^{QR^{\prime }})-\frac 12\varepsilon
^2(\sum_{km}\Phi _{km}^{\prime 2}/\lambda _{km}+\eta )+o(\varepsilon ^3),$
where $\eta $ is the item comes from the $\varepsilon ^2$ item in the
expansion of $\rho _\varepsilon ^{QR}$ on $\rho ^{QR}$, this item can be
omitted comparing with the main item $\sum_{km}\Phi _{km}^{\prime 2}/\lambda
_{km}$ for large $N$. Denote $\Omega _{km}=\left\langle km\right| ^{\prime
}\Omega \left| km\right\rangle ^{\prime }$, and make use of Eq.(\ref{wave11}%
), we have
\begin{eqnarray}
\sum_{km}\Phi _{km}^{\prime 2}/\lambda _{km} &=&\sum_{km}\Omega
_{km}^2/\lambda _{km}  \nonumber \\
&=&(n!)^2N^{-2n}(N+1)^{-2n}  \nonumber \\
&&\times \sum_{j=0}^n\binom nj^{2j}B^jA^{n-j}
\end{eqnarray}
where Eq.(\ref{wave5}) has been used, and $B=N_B(N_B+1)\cosh ^4r,$ $%
A=N_A(N_A+1)\sinh ^4r$. At the limit of $N\rightarrow \infty ,$we have $%
N^{\prime }\rightarrow N,$ the difference of CI between $\rho _\varepsilon $
and $\rho $ is
\begin{eqnarray}
\lim_{N\rightarrow \infty }[I_c(\rho _\varepsilon ,\mathcal{E})-I_c(\rho ,%
\mathcal{E})] &=&-\frac 12\varepsilon ^2\left( n!\right) ^2N^{-2n}  \nonumber
\\
\times [1-2^{-2n}\sum_{j=0}^n\binom nj^2] &<&0.  \label{wave12}
\end{eqnarray}

If we have a linear combination of above type perturbations, see $c_1\left|
\mu \right| ^{2n_1}+c_2\left| \mu \right| ^{2n_2}$ $(n_1<n_2)$, the
interference term $\sum_k\phi _{1k}\phi _{2k}/\lambda _k=(-1)^{n_2}Tr(\xi _1%
\frac{d^{n_2}\rho }{dN^{n_2}})=0,$ the similar zero interference can be
proved for the perturbation to the joint state $\rho ^{QR^{\prime }}$. Thus
each item act separately to the coherent information. So that at the
sufficiently large input energy and at the single use of the channel, the
first order non-Gaussian perturbation to the input thermal state will not
improve the conjectured capacity of the Gaussian quantum channel.

In the $n$ use of the channel with an input Gaussian state $\rho _n,$the
algebraic equations of the symplectic eigenvalues \cite{Holevo} are not
analytically solvable. Fortunately, when the input state is a two mode
squeezed thermal state, we can obtain analytical result. For the input state
of $\rho _2=S_2(r_2)\rho \otimes \rho S_2^{\dagger }(r_2)$ , where $S_2(r_2)$
is the two mode squeezing operator with real parameter $r_2$, the coherent
information is $I_c(\rho _2,\mathcal{E}^{\otimes 2})=2\max \{0,g(d_0-\frac
12)-g(d_1-\frac 12)-g(d_2-\frac 12)\},$ where in the expression of $d_i,$ $E$
should be substituted with $E/2,$ now $E$ is the total energy of the two
mode state; and $x=1/\cosh ^2r_2.$ The detail of the calculation will be
given elsewhere. The conclusion is that the maximum coherent information is
achieved by the product thermal state $\rho \otimes \rho $ when the input
state is two mode squeezed thermal state at sufficiently large input energy.
Also with Lagrange multiplier method, it is easy to prove that at
sufficiently large input energy, for all $n$ mode product thermal states,
the product of identical thermal state will achieve the maximal of coherent
information. Thus an unbalanced energy distribution among modes will not
increase the total coherent information. Hence, we have proved that for all
product Gaussian state inputs and all product of two mode squeezed thermal
state inputs, the maximum of the coherent information is achieved by product
identical thermal state $\rho _n=\rho ^{\otimes n}$ for sufficient large
input energy.

We now turn to the first order multi-mode perturbation. We will omit the
case that the perturbation can be treated separately for each mode. What
left is the perturbation with particle exchange among modes. A typical case
is $\chi _{n\varepsilon }(\mathbf{\mu })=\chi _n(\mathbf{\mu }%
)[1+\varepsilon (c\mu _1^{k_1}\mu _2^{k_2}\cdots \mu _n^{k_n}\mu
_1^{*l_1}\mu _2^{*l_2}\cdots \mu _n^{*l_n}+c^{*}\mu _1^{*k_1}\mu
_2^{*k_2}\cdots \mu _n^{*k_n}\mu _1^{l_1}\mu _2^{l_2}\cdots \mu _n^{l_n})]$
with $\sum_{i=1}^nk_i=\sum_{i=1}^nl_i=m$ (the requirement of first order
perturbation). We may denote the perturbation as $(\mathbf{k},\mathbf{l}),$
with vectors $\mathbf{k}=(k_1,k_2,\cdots ,k_n),\mathbf{l}=(l_1,l_2,\cdots
,l_n)$ and $\mathbf{k\neq l}$ $.$ The perturbed input state is $\rho
_{n\varepsilon }=\rho ^{\otimes n}+\phi .$ To simplify the calculation, we
introduce a generation function $I_\phi \left( \mathbf{\tau ,\sigma }\right)
=\int \left[ \Pi _i\frac{d^2\mu _i}\pi \right] \chi _n(\mathbf{\mu })D(-%
\mathbf{\mu })\exp [\mathbf{\mu \cdot \tau +\mu }^{*}\cdot \mathbf{\sigma }%
], $ then
\begin{equation}
\phi =\left( c\frac{\partial ^{2m}I_\phi \left( \mathbf{\tau ,\sigma }%
\right) }{\Pi _i(\partial \tau _i^{k_i}\partial \sigma _i^{l_i})}+c^{*}\frac{%
\partial ^{2m}I_\phi \left( \mathbf{\tau ,\sigma }\right) }{\Pi _i(\partial
\tau _i^{l_i}\partial \sigma _i^{k_i})}\right) _{\mathbf{\tau =\sigma =0}}.
\end{equation}
The eigenspace of $\rho _{n\varepsilon }$ can be classified as subspaces
according to the eigenvalue $\rho ^{\otimes n}$. When the eigenvalue of $%
\rho ^{\otimes n}$ is $\lambda _j=(1-v)^nv^j,$ the basis of the subspace can
be $\left| j_1,j_2,\cdots ,j_n\right\rangle =\left| \mathbf{j}\right\rangle $
, with $\sum_{i=1}^nj_i$ $=j.$ The action of $\phi $ will keep $j$
invariant, that is, it is an operator in $j$ subspace. In the subspace, we
may specify $\phi $ as $M_j.$ The eigenvalues of $M_j$ is supposed to be $%
\Lambda _{j\mathbf{i}}$, then $\sum_i$ $\Lambda _{j\mathbf{i}}^2=Tr(M_j^2)$ $%
=\sum_{\mathbf{j}}\left\langle \mathbf{j}\right| \phi \left| \mathbf{j}%
\right\rangle \left\langle \mathbf{j}\right| \phi \left| \mathbf{j}%
\right\rangle =$ $\sum_{\mathbf{jj}^{\prime }}\left\langle \mathbf{j}\right|
\phi \left| \mathbf{j}^{\prime }\right\rangle \left\langle \mathbf{j}%
^{\prime }\right| \phi \left| \mathbf{j}\right\rangle =$ $\sum_{\mathbf{j}%
}\left\langle \mathbf{j}\right| \phi ^2\left| \mathbf{j}\right\rangle ,$
where $\mathbf{j}^{\prime }$ may not be in the $j$ subspace. Here we have
used the fact that $\left\langle \mathbf{j}\right| \phi \left| \mathbf{j}%
^{\prime }\right\rangle =0$ for $\mathbf{j}^{\prime }\notin $the subspace of
$j.$ The perturbation to the entropy will be
\begin{eqnarray}
S(\rho _{n\varepsilon })-S(\rho ^{\otimes n}) &=&-\frac 12\varepsilon
^2\sum_{j,\mathbf{i}}\frac{\Lambda _{j\mathbf{i}}^2}{\lambda _j}%
+o(\varepsilon ^3)  \nonumber \\
&=&-\frac 12\varepsilon ^2Tr(\phi ^2/\rho ^{\otimes n})+o(\varepsilon ^3).
\end{eqnarray}
Where the linear term of $\varepsilon $ is nullified by the fact that $%
Tr(M_j)=0.$ For
\begin{equation}
Tr[I_\phi \left( \mathbf{\tau ,\sigma }\right) I_\phi \left( \mathbf{\tau }%
^{\prime }\mathbf{,\sigma }^{\prime }\right) /\rho ^{\otimes n}]=\exp [-%
\frac{\mathbf{\tau \cdot \sigma }^{\prime }}N-\frac{\mathbf{\tau }^{\prime }%
\mathbf{\cdot \sigma }}{N+1}],  \label{wave13}
\end{equation}
Thus
\begin{equation}
Tr(\phi ^2/\rho ^{\otimes n})=2\left| c\right| ^2\frac{\Pi _i(k_i!l_i!)}{%
[N(N+1)]^m}.
\end{equation}
In obtain Eq.(\ref{wave13}), we first work out the operator integral of $I$
part, which have an operator that conceal the $1/\rho ^{\otimes n}$
operator. Eq.(\ref{wave13}) exhibits that any interference item of $Tr(\phi
\phi ^{\prime }/\rho ^{\otimes n})$ type will be nullified for $(\mathbf{k},%
\mathbf{l})\neq $ $(\mathbf{k}^{\prime },\mathbf{l}^{\prime })$. Thus each
perturbation item contributes to the entropy separately.

The perturbation to the joint $QR$ state is more sophisticated. With almost
the same routine as we do in obtaining the perturbation operator in one mode
situation, we can get $\rho _{n\varepsilon }^{QR}=\rho ^{QR\otimes
n}+\varepsilon \Phi .$ Care should be taken in obtaining the operator $\Phi $
for complex coefficient $c,$ when $c$ is real, there are no problem; while $%
c $ is complex, there are extra phase factors in transforming the eigenbasis
of $\phi $ into the basis of direct product of the modes. However the phase
factor can be absorbed in the purification process, that is, if we have $%
e^{i\theta }\left| \mathbf{j}\right\rangle \left| \mathbf{j}\right\rangle $
in the purified state, it will make no difference with $\left| \mathbf{j}%
\right\rangle \left| \mathbf{j}\right\rangle $ in obtaining the reduced
state. We thus have $\Phi =\frac 12(\Phi _0+\Phi _0^{\dagger }),$ The
generation function of $\Phi _0$ is $I_{\Phi _0}=\exp \left[ \frac{\mathbf{%
\sigma \cdot (\tau -a}^{\dagger }\mathbf{)}}{N+1}\right] \exp \left[ \frac{%
\mathbf{\tau \cdot a}}N\right] \rho ^{QR\otimes n}.$ The action of the
channel then is $I_{\Phi _0}^{\prime }=$ $(\mathcal{E}\otimes \mathbf{I)}%
I_{\Phi _0}=\exp (p\mathbf{\tau \cdot b}^{\dagger })\exp [\frac{\mathbf{\tau
\cdot \sigma }}{N+1}-p\mathbf{\sigma \cdot b]}\rho ^{QR^{\prime }\otimes n}$
according to lemma 1 and lemma2, where $p=[N(N+1)]^{-1/2}.$ The contribution
to the entropy should be evaluated in the eigenbasis of $\rho ^{QR^{\prime
}\otimes n}$. We may denote the subspace of $\rho ^{QR^{\prime }\otimes n}$as%
$\left| i,\mathbf{i};j,\mathbf{j}\right\rangle ^{\prime }$ which has
eigenvalue $\lambda _{ij}=(1-v_A)^n(1-v_B)^nv_A^iv_B^j$ $.$ In this
subspace, we denote $\Phi _0^{\prime }$ as $M_{ij},$ the elements of $M_{ij}$
are $\left\langle i,\mathbf{i},j,\mathbf{j}\right| ^{\prime }\Phi _0^{\prime
}\left| i,\mathbf{i}^{\prime };j,\mathbf{j}^{\prime }\right\rangle ^{\prime
},$ the sum of the square of the eigenvalue of is $TrM_{ij}^2$. We obtain
the contribution to the entropy by first evaluating $\left\langle i,\mathbf{%
i;}j,\mathbf{j}\right| ^{\prime }I_{\Phi _0}^{\prime }\left| i,\mathbf{i}%
^{\prime }\mathbf{;}j,\mathbf{j}^{\prime }\right\rangle ^{\prime
}=\left\langle i,\mathbf{i;}j,\mathbf{j}\right| V^{\dagger \otimes n}I_{\Phi
_0}^{\prime }V^{\otimes n}\left| i,\mathbf{i}^{\prime }\mathbf{;}j,\mathbf{j}%
^{\prime }\right\rangle ,$ which is $\exp [\mathbf{\tau \cdot \sigma /(}%
N+1)] $ $\left\langle i,\mathbf{i;}j,\mathbf{j}\right| \exp [p\mathbf{\tau
\cdot (b}^{\dagger }\cosh r+\mathbf{a}\sinh r)]$ $\exp [-p\mathbf{\sigma
\cdot (b}\cosh r+\mathbf{a}^{\dagger }\sinh r)\left| i,\mathbf{i}^{\prime
};j,\mathbf{j}^{\prime }\right\rangle .$ We expand the exponent of the
operators to drop the terms that do not keep the total particle numbers in $%
A $ and $B$ parts respectively. Denote $I_B(\mathbf{\tau ,\sigma }%
)=\sum_{k=0}^\infty \frac{(-1)^k}{k!^2}(p\cosh r)^{2k}(\mathbf{\tau \cdot b}%
^{\dagger })^k(\mathbf{\sigma \cdot b})^k,$ $I_A(\mathbf{\tau ,\sigma }%
)=\sum_{k=0}^\infty \frac{(-1)^k}{k!^2}(p\sinh r)^{2k}(\mathbf{\tau \cdot a}%
)^k(\mathbf{\sigma \cdot a}^{\dagger })^k.$ Then in the calculation of the
contribution to the entropy become a trace on the whole space, the
restriction on the subspace is removed. We have the generation function
\begin{eqnarray}
F &=&\exp [(\mathbf{\tau \cdot \sigma +\tau }^{\prime }\mathbf{\cdot \sigma }%
^{\prime }\mathbf{)/(}N+1)]  \nonumber \\
&&\times Tr[(I_B(\mathbf{\tau ,\sigma })I_B(\mathbf{\tau }^{\prime }\mathbf{%
,\sigma }^{\prime })/\rho _B^{\otimes n}]  \nonumber \\
&&\times Tr[I_A(\mathbf{\tau ,\sigma })I_A(\mathbf{\tau }^{\prime }\mathbf{%
,\sigma }^{\prime })/\rho _A^{\otimes n}]  \nonumber \\
&=&\sum_{m=0}^\infty \sum_{l=0}^m\frac{B^lA^{m-l}}{[l!(m-l)!]^2[N(N+1)]^{2m}}
\nonumber \\
&&\times (\mathbf{\tau \cdot \sigma }^{\prime })^m(\mathbf{\tau }^{\prime }%
\mathbf{\cdot \sigma })^m.  \label{wave14}
\end{eqnarray}
, with which the perturbation to the entropy can be calculated, where we
have used another generation function in evaluating $F,$ and at the final
step we exchange the orders of summation and make use of Eq.(\ref{wave5}) to
conceal the factor $\exp [(\mathbf{\tau \cdot \sigma +\tau }^{\prime }%
\mathbf{\cdot \sigma }^{\prime }\mathbf{)/(}N+1)].$ The generation function
is about that the two ingredients both come from $\Phi _0,$ if both come
from $\Phi _0^{\dagger },$the result will be the same. In the case
intercross of $\Phi _0$ and $\Phi _0^{\dagger },$ we should substitute $(%
\mathbf{\tau \cdot \sigma }^{\prime })^m(\mathbf{\tau }^{\prime }\mathbf{%
\cdot \sigma })^m$ in Eq.(\ref{wave14}) by $(\mathbf{\tau \cdot \tau }%
^{\prime })^m(\mathbf{\sigma }^{\prime }\mathbf{\cdot \sigma })^m$.

\begin{eqnarray}
\sum_{ij}\frac{Tr[\frac 12(M_{ij}+M_{ij}^{\dagger })]^2}{\lambda _{ij}}
&=&2\left| c\right| ^2\frac{\Pi _i(k_i!l_i!)}{[N(N+1)]^{2m}}  \nonumber \\
&&\times \sum_{l=0}^m\binom ml^2B^lA^{m-l}.
\end{eqnarray}
The situation is strictly like that of the single mode case. Eq.(\ref{wave14}%
) indicates that each perturbation term contributes to the entropy
separately. Also, the state $\rho _\varepsilon ^{QR\otimes n}$ then is
expanded in $\varepsilon $ to the linear item (the $\varepsilon ^2$ term is
less important in the large $N$ limit).

We have shown that all first order perturbation to the input product
identical thermal state can only decrease the coherent information at large
input energy. In one mode case of Gaussian input, in two mode of squeezed
thermal input, and non-Gaussian perturbation of one mode as well multi-mode
to thermal state input, thermal or their identical product achieve maximum
of the coherent information at infinitive input energy. For all inputs of
product Gaussian states, product two-mode squeezed thermal states, we have
proved that the channel capacity of additive Gaussian quantum channel is
described by conjectured formula (\ref{wave1}) and achieved by product of
identical thermal state. All kinds of the first order perturbations
(non-Gaussian) to the product of identical thermal state input will lead to
a less coherent information for sufficient large input energy. Whether a
bigger coherent information can be achieved by a state which is far from the
product of identical thermal state still remains open.

\textit{\ }Funding by the National Natural Science Foundation of China
(Grant No. 10575092), Zhejiang Province Natural Science Foundation (Grant
No. RC104265) are gratefully acknowledged.

\end{document}